\documentclass{PoS}
\usepackage{mathrsfs}

\title{Toward a well defined monopole creation operator}

\ShortTitle{Monopole creation operator}

\author{\speaker{Claudio Bonati}\\
        Dipartimento di Fisica, Universit\`a di Pisa and INFN, Largo Pontecorvo 3, I-56127 Pisa, Italy.\\
        E-mail: \email{claudio.bonati@pi.infn.it}}

\author{Guido Cossu\\
        KEK Theory Center, 1-1 Oho, Tsukuba-shi, Ibaraki 305-0801, Japan. \\
        E-mail: \email{cossu@post.kek.jp}}

\author{Massimo D'Elia\\
        Dipartimento di Fisica, Universit\`a di Genova and INFN, Via Dodecaneso 33, 16146 Genova, Italy.\\
        E-mail: \email{massimo.delia@ge.infn.it}}

\author{Adriano Di Giacomo\\
        Dipartimento di Fisica, Universit\`a di Pisa and INFN, Largo Pontecorvo 3, I-56127 Pisa, Italy.\\
        E-mail: \email{digiaco@df.unipi.it}}

\abstract{The lattice implementation of monopole creation operator proves to have problems related to bulk transitions that can possibly 
affect the interpretation of its mean value as an order parameter for monopole condensation. Preliminary evidence is presented that these 
unexpected behaviours are in fact only due to lattice artefacts and do not spoil the physical interpretation of the monopole operator.}

\FullConference{The XXVIII International Symposium on Lattice Filed Theory\\
		 June 14-19,2010\\
		 Villasimius, Sardinia Italy}

\newcommand{\eg}{\emph{e.g.} }
\newcommand{\ie}{\emph{i.e.} }
\newcommand{\vev}{\emph{vev} }
\newcommand{\Tr}{\ensuremath{\mathrm{Tr}}}
\newcommand{\ds}{\ensuremath{\displaystyle}}
\newcommand{\mbf}[1]{\ensuremath{\mathbf{#1}}}
\newcommand{\dx}[2][x]{\ensuremath{\mathrm{d}^{#2}\mathbf{#1}}}

\makeatletter
\@ifundefined{[}{\newcommand{\[}{\ensuremath{\lbrack}} }{\renewcommand{\[}{\ensuremath{\lbrack}} }
\@ifundefined{]}{\newcommand{\]}{\ensuremath{\rbrack}} }{\renewcommand{\]}{\ensuremath{\rbrack}} }
\makeatother
\DeclareRobustCommand{\eqref}[1]{Eq.~(\ref{#1})}

\begin{document}

\section{Introduction}

The idea that color confinement can be produced by dual superconductivity of the vacuum was introduced in 
Refs.~\cite{mandelstam, thooft77} and since then it attracted a big interest. However, despite its intuitive nature, 
this proposal is not easy to demonstrate, the main difficulty being that it is far from trivial to identify the magnetic 
degrees of freedom. 

In Ref.~\cite{thooft81} the idea was advocated that any operator in the adjoint representation of the gauge group 
can be used as an effective Higgs field to identify the magnetic $U(1)$ group (abelian projection), physics being 
independent of that choice. This revealed to be true as far as monopole condensation is concerned while for 
monopole detection a strong dependence on the abelian projection adopted is observed (see the discussion in \cite{bdlp}). 

The correct procedure to detect monopole condensation, that is Higgs breaking of the $U(1)$ magnetic 
symmetry, is to evaluate the vacuum expectation value (\emph{vev}) of a magnetically charged operator $\mu$: this \vev 
has to be zero in the normal phase and can be different from zero in the dual superconducting phase. 
In order to pursue this strategy a magnetically charged operator $\mu$ was introduced and developed in Refs.~\cite{ddp, ddpp, dp, dlmp,3}.
Similar constructions were developed in Refs.~\cite{vpc, ccos}.

The operator $\mu$, however, proved to a more rigorous analysis to have a bad infrared behaviour, \ie a bad thermodynamical limit,
as shown in Refs.~\cite{unpub, gl}.
 
In the following sections we will sketch the construction of the $\mu$ operator and present a possible way to overcome
these problems.

\section{Construction of $\mu$}

The simplest way to construct a magnetically charged operator is to define it as the operator that adds a monopole to the state
to which it is applied (\cite{mss}). This is easily done in the Schr\"{o}dinger representation: if we denote by $\Pi_{i}(\mbf{x})$
the canonical momenta conjugate to the transverse components of the gauge fields $A_{\mu}(\mbf{x})$, the operator
$\mu$ defined by 
\begin{equation}\label{mu}
\mu(\mbf{y})=\exp\left( i\int \dx{3}\; b_i(\mbf{x},\mbf{y})\Pi_i(\mbf{x})\right)
\end{equation}
is the translation operator of the gauge field by $\mbf{b}(\mbf{x},\mbf{y})$:
\begin{equation}
\mu(\mbf{y})|\mbf{A}(\mbf{x})\rangle=|\mbf{A}(\mbf{x})+\mbf{b}(\mbf{x},\mbf{y})\rangle
\end{equation}
This is equivalent to say that the following commutation relations hold
\begin{eqnarray}
&& \[ A_i(\mbf{x}),\mu(\mbf{y})\]=b_i(\mbf{x},\mbf{y})\mu(\mbf{y}) \label{cc}\\
&& \[ \Pi_i(\mbf{x}), \mu(\mbf{y})\]=0
\end{eqnarray}
Because of the linearity with respect to $A_{\mu}$ of the 't Hooft tensor, it follows from \eqref{cc} that
\begin{equation}
\[Q, \mu(\mbf{y})\]=m\;\mu(\mbf{y})
\end{equation} 
where $Q$ is the magnetic charge operator and $m$ is the charge of the field $\mbf{b}(\mbf{x},\mbf{y})$. 
The operator $\mu$ is thus charged if $\mbf{b}(\mbf{x},\mbf{y})$ is the field in $\mbf{x}$ of a monopole located in 
$\mbf{y}$.  

On the lattice the canonical momenta correspond (with the Wilson action) to the mixed spatial-temporal plaquettes and 
the operator in \eqref{mu} can be rewritten (up to $O(a^2)$ lattice artefacts) as
\begin{equation} \label{latticemu}
\mu=\exp(-\beta\Delta S)\qquad 
\Delta S=\sum_{\mbf{n}}\Tr\{\Pi_{i0}(\mbf{n}, t)-\Pi'_{i0}(\mbf{n}, t)\}
\end{equation}
where
\begin{eqnarray}
&& \Pi_{i0}(\mbf{n}, t) = U_{i}(\mbf{n}, t)U_{0}(\mbf{n}+\mbf{\hat{\imath}}, t)
U_i(\mbf{n}, t+1)^{\dag} U_0(\mbf{n}, t)^{\dag}\\
&& \Pi'_{i0}(\mbf{n}, t)=U_i(\mbf{n}, t) U_0(\mbf{n}+\mbf{\hat{\imath}}, t) M_i(\mbf{n}+\mbf{\hat{\imath}}, t)
U_i(\mbf{r}, t+1)^{\dag}U_0(\mbf{n}, t)^{\dag}\\
&& M_j(\mbf{n},t)=\left\{ \begin{array}{lll}
    \exp\left(i\; b_j(\mbf{n},\mbf{y})\; \hat{\Phi} \right) & \mathrm{if} & t=0 \\
    0 & \mathrm{if} & t\ne 0
    \end{array}\right. \label{randproj}
\end{eqnarray}
and $\hat{\Phi}$ is the generator of the gauge group which identifies the magnetic $U(1)$.  
The operator of \eqref{randproj} corresponds to the choice of the so-called random abelian projection, which simplifies 
the numerical calculations (\cite{3}). The fact that
the operator in \eqref{latticemu} adds a monopole to a given configuration can be shown directly by an explicit gauge 
transformation (see Refs.~\cite{dp, dlmp}).  If we denote by $S$ the Wilson action, the expectation value of $\mu$ is  
given by  
\begin{equation}\label{meanmu}
\langle \mu\rangle=\frac{\int \mathscr{D}U\; e^{-\beta(S+\Delta S)}}{\int \mathscr{D}U\; e^{-\beta S}}
\end{equation}

\begin{figure}[b]
\centering
\includegraphics[width=0.6\textwidth,clip]{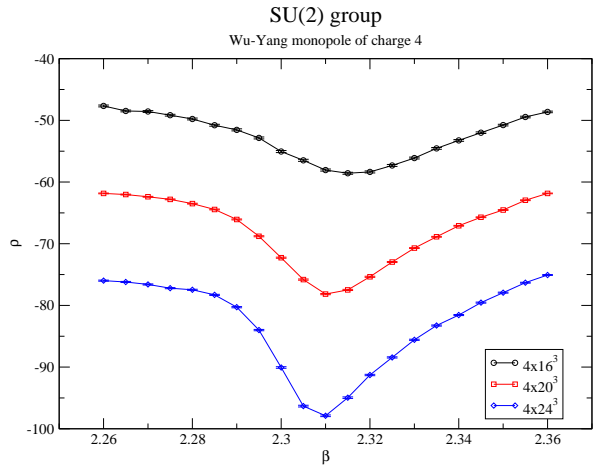}
\caption{Value of the $\rho$ observable defined in \eqref{rho}.}
\label{rho_fig}
\end{figure}

For computational reasons it is convenient to use, instead of $\langle \mu\rangle$, its logarithmic derivative
\begin{equation}\label{rho}
\rho=\frac{\mathrm{d}}{\mathrm{d}\beta}\log\langle\mu\rangle=\langle S\rangle_S-\langle S+\Delta S\rangle_{S+\Delta S}
\end{equation}
Indeed, since $\langle\mu\rangle=1$ at $\beta=0$, the $\mu$ \vev can be reconstructed as $\langle\mu\rangle=\exp\left(\int_0^{\beta}\rho(x)
\mathrm{d}x\right)$.

The quantity $\rho$, which is the one measured in lattice simulations, should have a finite value in the thermodynamical limit for temperatures 
smaller than the deconfinement temperature ($T<T_c$) and should develop a negative peak at $T_c$, scaling with the appropriate
critical indices (see \eg Ref.~\cite{dlmp}). For $T>T_c$, if the dual superconductor picture is correct, 
$\langle\mu\rangle=0$ in the thermodynamical limit and $\rho\to -\infty$.  

The first simulations were done at $N_t=4$ and different spatial extensions $N_s$ and seemed to agree with these expectations (\cite{dlmp,3}). A more
careful analysis performed with different values of $N_t$ \cite{unpub} showed that 
\begin{enumerate}
\item the observed peak did not move by varying $N_t$ as expected from the known dependence of $T$ on $\beta$
\item for $T<T_c$ the values of $\rho$, instead of tending to a finite limit as $N_s\to\infty$, keep growing negative with the volume, so that
$\langle\mu\rangle=0$ also for $T<T_c$
\item a physical peak appears at the right position, superposed to the background of 1. and 2.
\end{enumerate}
The direct meaning of $\mu$ as an order parameter was thus spoiled.

For abelian theories the operator in \eqref{latticemu} clearly shows the correct behaviour for $T<T_c$. Moreover  it was shown 
in Ref.~\cite{dp} to be equivalent to the one constructed in Ref.~\cite{fm} by explicitly using the duality transformation.
For non-abelian theories this is less clear: it is by now known (\cite{unpub, gl}) that the quantity $\rho$ defined in \eqref{rho} 
seems to diverge also for $T<T_c$. 

\begin{figure}
\centering
\includegraphics[width=0.6\textwidth,clip]{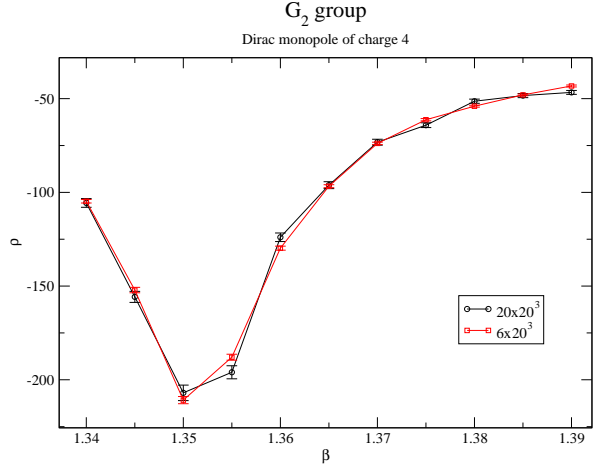}
\caption{Bulk transition in $G_2$ gauge theory}
\label{g2_bulk_fig}
\end{figure}

This behaviour is shown in Fig.~\ref{rho_fig} for case of the group $SU(2)$ on lattices of temporal extent $N_t=4$.
The deconfinement transition is located, in the thermodynamical limit, at $\beta_c=2.2986(6)$ and, as expected, $\rho$
develops a negative peak in the neighbourhood of this $\beta$ value (finite size corrections are obviously to be
expected). However it looks like there is a strong volume dependent background also for $T<T_c$, where $\rho$ should 
have a well defined thermodynamical limit.

A plausible working hypothesis was that the source of this problem were the $O(a^2)$ lattice artefacts which are present in the definition 
\eqref{latticemu}. If this were true it seems more appropriate to use the Wu-Yang form of the abelian monopole for the field 
$\mbf{b}(\mbf{x},\mbf{y})$ instead of the Dirac form used in the past. This is because in the Dirac expression
\begin{equation}\label{dirac_mono}
\mbf{b}=-g\frac{(1+\cos\theta)}{r\sin\theta}\mbf{e}_{\phi}
\end{equation}
the field is large near the $\hat{z}$ axis also far away from the monopole, while in the Wu-Yang formulation
\begin{equation}\label{wuyang_mono}
\left\{ \begin{array}{ll}
\mbf{b}^N=g\frac{\ds (1-\cos\theta)}{\ds r\sin\theta}\,\mbf{e}_{\phi} & \hspace{1cm} 0\le\theta<\pi/2+\epsilon\\
\rule{0mm}{7mm}
\mbf{b}^S=-g\frac{\ds  (1+\cos\theta)}{\ds r\sin\theta}\,\mbf{e}_{\phi} & \hspace{1cm} \pi/2-\epsilon<\theta\le \pi 
\end{array} \right.
\end{equation}
the only singularity of the field is at the monopole location. Indeed by using the field \eqref{wuyang_mono} the background
is reduced typically by about a factor of $3$. The general behaviour is nevertheless divergent in the thermodynamical limit, 
see Fig.~\ref{rho_fig}.

\begin{figure}
\centering
\includegraphics[width=0.6\textwidth,clip]{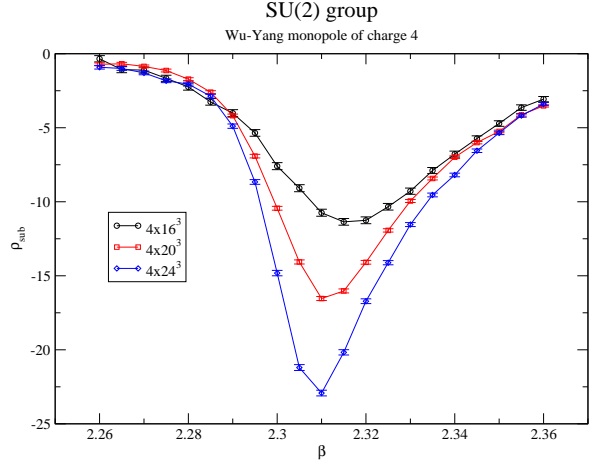}
\caption{Value of the $\rho_{\rm sub}$ observable defined in \eqref{rhosub}.}
\label{rhosub_fig}
\end{figure}

\section{A possible solution}

If the divergence of $\rho$ below $\beta_c$ is driven by lattice artefacts, it comes from short distances and therefore it 
has to be independent of the temperature, 
that is of the temporal extent of the lattice. This proves indeed to be the case, both with the Wu-Yang and with the Dirac monopole.

As an example we show in Fig.~\ref{g2_bulk_fig} the values of $\rho$ calculated for the $G_2$ group on the lattices $6\times 20^3$ 
and $20\times 20^3$. The (first order) deconfinement transition of the model for $N_t=6$ is located between 
$\beta=1.39$ and $\beta=1.4$ (see Ref.~\cite{g2, g2_bis}), while the big bump near $\beta=1.355$ is just the non-physical 
bulk transition (in fact it is just an analytic crossover). As expected for a purely UV effects the value of $\rho$ 
does not depend on the temporal extent of the lattice.

This suggests a possible method to obtain a reliable definition of $\rho$: subtract from the $\rho$ values calculated on 
asymmetrical lattices (finite temperature) the values calculated at zero temperature:
\begin{equation}\label{rhosub}
\rho_{\rm sub}=\rho-\rho_{(T=0)}
\end{equation}
In terms of the $\mu$ operator this amounts to use $\langle\mu\rangle/\langle\mu\rangle_{T=0}$ instead of just 
$\langle\mu\rangle$. 

The results obtained by using the subtracted form \eqref{rhosub} are shown in Fig.~\ref{rhosub_fig} and now the behaviour 
for $\beta<\beta_c$ is the correct one. Near the deconfinement transition it can be shown (see \eg Ref.~\cite{dlmp}) that 
the minimum of the dip in the $\rho$ values scales as $N_s^{1/\nu}$, where $N_s$ is the lattice spatial extent, while
its width varies according to the scaling law $N_s^{-1/\nu}$. 

\begin{figure}
\centering
\includegraphics[width=0.6\textwidth,clip]{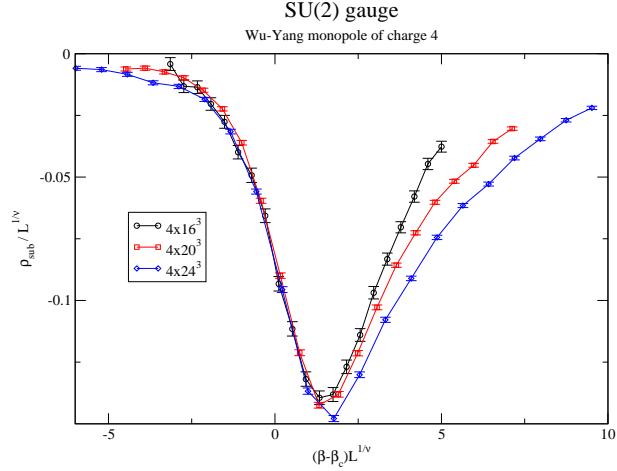}
\caption{Rescaled values of $\rho_{\rm sub}$.}
\label{rhosubscaled_fig}
\end{figure}

These scaling laws are shown to be satisfied in Fig.~\ref{rhosubscaled_fig}. The curves are drown by use of the known value of the
deconfinement coupling, $\beta_c=2.2986(6)$, and $\nu=0.6301(4)$, which is the value of the $\nu$ critical index for the 
3d Ising universality class. For $\beta<\beta_c$ the scaling is very good near the transition, while for $\beta>\beta_c$ 
data are consistent with the expectation that $\rho_{\rm sub}\to -\infty$ as $V\to\infty$. A careful analysis is on the way.

\section{Conclusions}

We have presented indications that the unexpected behaviour of the order parameter 
$\langle\mu\rangle$ for monopole condensation  in non-abelian gauge theories has to be ascribed to lattice artefacts. In order
to disentangle these artefacts from the physical signal one possibility is to renormalize the expectation value of the
$\mu$ operator by its zero temperature value. This procedure is shown to give reliable results, which nicely 
agree with the predictions based on universality arguments. Further studies on this line are on the way.

\end{document}